\documentclass[letterpaper,11pt]{extarticle}
\usepackage{import}
\usepackage{local}
\usepackage[left=.8in,top=.8in,bottom=.8in,right=.8in]{geometry}

\usepackage{adjustbox}
\usepackage{algorithm}
\usepackage{algpseudocode}
\usepackage{amsmath}
\usepackage[numbers]{natbib}

\usepackage[version=4]{mhchem}

\usepackage{tikz}
\usetikzlibrary{quantikz2}

\usepackage{siunitx}
\DeclareSIUnit\hartree{\text {\ensuremath {E}}_{\mathrm {h}}}
\DeclareSIUnit\angstrom{\text {Å}}

\algdef{SE}[DOWHILE]{Do}{doWhile}{\algorithmicdo}[1]{\algorithmicwhile\ #1}

\date{\today}

\begin{document}

\title{Iterative Qubit Coupled Cluster using only Clifford circuits}

%\author{James Brown}
%\affiliation{qBraid Co., Chicago, IL 60615, USA}
%\altaffiliation{The work of J.B. was performed before joining qBraid.}
%\author{Marc P. Coons}
%\affiliation{Dow, Core R\&D, Chemical Science, 1776 Building, Midland, MI, 48674, USA}
%\author{Erika Lloyd}
%\author{Alexandre Fleury}
%\affiliation{SandboxAQ, Palo Alto, CA 94301, USA}
%\author{Krzysztof Bieniasz}
%\affiliation{German Aerospace Centre (DLR), Institute for AI Safety and Security, 89081 Ulm, Germany}
%\altaffiliation{The work of K.B. was performed before joining the German Aerospace Centre.}
%\author{Valentin Senicourt}
%\author{Arman Zaribafiyan}
%\affiliation{SandboxAQ, Palo Alto, CA 94301, USA}

\author{%
\textbf{James Brown\textcolor{Accent}{\textsuperscript{3}}, %
Marc P. Coons\textcolor{Accent}{\textsuperscript{2}}, %
Erika Lloyd\textcolor{Accent}{\textsuperscript{1}}, %
Alexandre Fleury\textcolor{Accent}{\textsuperscript{1,*}}, %
Krzysztof Bieniasz\textcolor{Accent}{\textsuperscript{4}}, %
Valentin Senicourt\textcolor{Accent}{\textsuperscript{1}}, %
Arman Zaribafiyan\textcolor{Accent}{\textsuperscript{1}} }\\
\begin{small}\textcolor{Accent}{\textsuperscript{1}}SandboxAQ, Palo Alto, CA 94301, USA \\ 
\textcolor{Accent}{\textsuperscript{2}}Dow, Core R\&D, Chemical Science, 1776 Building, Midland, MI, 48674, USA \\ 
\textcolor{Accent}{\textsuperscript{3}}qBraid Co., Chicago, IL 60615, USA \\ 
\textcolor{Accent}{\textsuperscript{4}}German Aerospace Centre (DLR), Institute for AI Safety and Security, 89081 Ulm, Germany \\ 
\textcolor{Accent}{\textsuperscript{*}}Corresponding author: \textcolor{Accent}{alexandre.fleury@sandboxaq.com} \\ 
The work of J.B. was performed before joining qBraid.\\
The work of K.B. was performed before joining the German Aerospace Centre. \\\end{small}
}

\maketitle

\begin{doublespacing}
%\begin{linenumbers}

\noindent

\begin{abstract}
The performance of quantum algorithms for ground-state energy estimation is directly impacted by the quality of the initial state, where quality is traditionally defined in terms of the overlap of the input state with the target state. An ideal state preparation protocol can be characterized by being easily generated classically and can be transformed to a quantum circuit with minimal overhead while having a significant overlap with the targeted eigenstate of a given Hamiltonian. We propose a method that meets these requirements by introducing a variant of the iterative qubit coupled cluster (iQCC) approach, which exclusively uses Clifford circuits. These circuits can be efficiently simulated on a classical computer, with polynomial scaling according to the Gottesman-Knill theorem.
Since the iQCC method has been developed as a quantum algorithm firstly, our variant can be mapped naturally to quantum hardware. We additionally implemented several optimizations to the algorithm enhancing its scalability. We demonstrate the algorithm's correctness in ground-state simulations for small molecules such as \ce{H2}, \ce{LiH}, and \ce{H2O}, and extend our study to complex systems like the titanium-based compound \ce{Ti(C5H5)(CH3)3} with a (20, 20) active space, requiring 40 qubits. Results show that the convergence of the algorithm is well-behaved, and the ground state can be represented accurately. Moreover, we show an automated workflow for restricting the qubit active space, thus relieving computational resources by considering only qubits affected by non-trivial operations.
\end{abstract}

\section{Introduction}\label{sec:intro}

% Justify state preparation research.
A special interest for the application of quantum computers is the computation of ground state energies in many-body physics. Many methods have been proposed~\cite{kitaev1995quantum,stair2020multireference,cortes2022quantum,low2024quantum} that share the quality of the initial state as a common component for success. Given these diverse approaches, and the variety of chemical systems, the best combination of quantum algorithm and initial state preparation is likely system-dependent. Ultimately, the speedup gained by an energy estimation algorithm must be greater than the overhead of producing the initial state. For this reason, the Hartree-Fock method is often used due to its computational efficiency. Unfortunately, the state preparation quality, measured via the overlap with the targeted state, is insufficient for large and strongly correlated systems~\cite{kohn1999nobel}, which are potential candidates for applications of quantum algorithms. These factors motivate focusing research efforts toward improving initial state preparation with shallow circuits and low classical overhead. 

% Summary of works around initial state preparation.
Approaches to initial state preparation seek a balance between classical and quantum computing resources. Proposed methods include adiabatic state preparation, which consist of evolving a quantum state with a time-dependent operator that begins as a Fock operator, and changes slowly to a fully correlated Hamiltonian~\cite{sugisaki2022adiabatic,nishiya2024first,hejazi2023adiabatic}. More recent methods encode classical states onto a quantum computer, but although they seem promising, they still retain the overhead of encoding sum of Slater determinants (SOS) or Matrix Product States (MPS) with limited applicability to particular systems~\cite{fomichev2023initial}. Quantum resource efficient methods use variational quantum algorithms (VQAs)~\cite{QCC,iQCC,ADAPT-VQE,VQE,Tang_2021,QEB} to produce shallow circuits at the cost of performing many measurements for energy evaluations. There has been work to mitigate this measurement problem by using Clifford pre-optimization~\cite{sun2024toward,schleich2023partitioning,mishmash2023hierarchical}, but there is room for further optimization by addressing the measurement problem more strategically.

% Description of VQAs.
In VQAs, a parameterized ansatz is chosen to encode the state of a given system which can be mapped to one or more eigenvectors of a qubit Hamiltonian (Eq.~\ref{eq:qubit_hamiltonian}).

\begin{equation}\label{eq:qubit_hamiltonian}
    \hat{H} = \sum_{i} h_i \mathcal{P}_i = \sum_i h_i \bigotimes_{j=1}^{n_{\text{qubits}}} p_j
\end{equation}

where $p_j \in \{ I, \sigma_x, \sigma_y, \sigma_z \}$, and $h_i$ is an expansion coefficient. $\mathcal{P}_i$ is therefore a tensor product of Pauli and identity operators, and is commonly referred to as a "Pauli word" operator. Also, the Pauli operators $\{\sigma_x, \sigma_y, \sigma_z \}$ are commonly abbreviated by $\{X, Y, X\}$, and this notation is used hereafter. For fermionic systems, the set of coefficients $h_i$ are derived from the one- and two-electron integrals of a second-quantized Hamiltonian using any of the popular mappings such as Jordan-Wigner (JW)~\cite{jordanUber1928}, Bravyi-Kitaev (BK)~\cite{bravyi2002fermionic}, and JKMN~\cite{JKMN}, also known as ternary-tree mapping. A qubit Hamiltonian expectation value is determined on a quantum computer by first preparing an initial qubit state according to the chosen ansatz and current parameter set, applying each $\mathcal{P}_i$ term individually or as groups of mutually commuting $\mathcal{P}_i$ terms via single-qubit rotations, and measuring the final qubit state in the computational basis. Minimization of this expectation value with respect to the ansatz parameters is performed on a classical computer and affords the optimal energy and parameter set for the selected ansatz. An effective ansatz should steer towards the optimal solution with short-depth circuit, and a minimal number of parameters.
 
% Measurement problem and clifford pre-opt.
The short-depth quantum circuits afforded by VQAs incurs a cost of an intractable number of measurements~\cite{gonthier2022measurements}. The number of measurements for an energy minimization is approximated by $D\approx M\lambda \theta s$, where $M$ is the number of distinct groupings of Pauli operators to measure, $\lambda$ is the number of iterations in the optimization procedure, $\theta$ is the number of parameters, and $s$ is the number of ``shots" (state preparation + measurement of the qubit register) used to sample an observable on a quantum computer at each iteration of the algorithm, for a given accuracy. Iterative ansatz building methods such as iterative qubit coupled cluster (iQCC) seek to reduce the number of variational parameters to optimize at a cost of increasing $M$~\cite{iQCC}. This is done through an explicit choice of expressing electron correlation through canonical transformation of the Hamiltonian instead of a circuit. Other methods attempt to reduce the number of iterations $\lambda$ by initializing the ansatz with better parameters, such as those obtained from Clifford pre-optimization~\cite{cheng2022clifford}. If a circuit is constructed using only Clifford gates, then a Clifford simulator can be used to measure an expectation value classically, in which case the eigenvalue can be computed in polynomial time using classical computers~\cite{aaronson2004improved}.
 
% High-level overview of the paper.
We highlight iQCC as a method for Clifford pre-optimization, and demonstrate its efficiency by simulating ground-state preparation of \ce{H2}, \ce{LiH}, \ce{H2O} and \ce{Ti(C5H5)(CH3)3}, where the last system is encoded by 40 qubits in our calculation~\cite{iQCC}. Our approach could scale further by by using a sparse wave function representation and multiple threads as was done in Ref.~\citenum{steiger2024sparse}. In Sections~\ref{sec:qcc}-\ref{sec:clifford_iqcc}, the algorithm fundamentals are discussed, while its optimization are detailed in Section~\ref{sec:scaling_up}. Finally, results of modeling the ground state of several molecular systems are included in Section~\ref{sec:results}.

\section{Qubit Coupled Cluster and its Iterative Version}\label{sec:qcc}

% Introducing QCC ansatz.
A promising wave function ansatz for VQA applications is achieved through the qubit coupled cluster (QCC) formalism~\cite{QCC, iQCC}. This method utilizes a parameterized ansatz $\hat{U}_{\rm{QCC}}$ that is expressed as a product of exponentiated Pauli words, 

\begin{equation}
\hat{U}_{\rm{QCC}} = \prod^{M}_{j=1}{{\rm{exp}} \left(-\frac{i}{2} \tau_{j} \mathcal{P}_j \right)}
\end{equation}

where $M$ is the number of Pauli word operators used in the ansatz, and $\tau_{j} \in [0, 2\pi[$. It is noted that the set of $\mathcal{P}_j$ operators appearing in $\hat{U}_{\rm{QCC}}$ is distinct from the set of $P_i$ operators present in Eq.~\ref{eq:qubit_hamiltonian}. For a system of $N$ qubits, the number of candidate $\mathcal{P}_j$ operators to consider for the QCC ansatz is $N_{\mathcal{P}} \leq 4^N$, and utilization of all candidates in $\hat{U}_{\rm{QCC}}$ provides a formally exact representation of a wave function for a molecular Hamiltonian. It was recognized in Ref.~\citenum{iQCC} that Pauli $Z$ operators do not contribute to variational energy lowering. Thus, the maximum cardinality for the set of Pauli operators is reduced to $N_{\mathcal{P}} \leq 3^N$.

% Details about getting the QCC generators.
Interestingly, the choice of mapping influences the number of candidate operators to consider for the QCC ansatz. In previous studies of QCC-based methods~\cite{QCC,iQCC}, restricting the rank of candidate operators, which is denoted as $\lvert \mathcal{P}_j \rvert$ and is a count of the number of Pauli operators, was proposed as a route to  systematically control the problem complexity and accuracy of the solution. Ryabinkin~et~al. noted that the rank of candidate operators respects $2 \leq \lvert \mathcal{P}_j\rvert \leq N$ when the JW and BK mappings are considered~\cite{QCC,iQCC}. It was also noted in Ref.~\citenum{iQCC}, and later by Tang~et~al.~\cite{Tang_2021}, that candidate Pauli words require an odd number of Pauli $Y$ operators. Piecing together the observations discussed above affords a lower bound for the pool size of candidate Pauli words to consider for the QCC ansatz as

\begin{equation}\label{eq.n_pool_pwgens}
  N_{\mathcal{P}} = \sum^{N}_{\substack{n_y=1\\n_y{\rm{odd}}}}{\sum^{N-n_y}_{n_x=n_0}{{\binom{N}{n_y}}{\binom{N-n_y}{n_x}}}}\;.
\end{equation}

In Eq.~\ref{eq.n_pool_pwgens}, $n_{x}$ and $n_{y}$ are the number of Pauli $X$ and $Y$ operators present in the candidate operator. The set of energy-lowering Pauli word candidates obtained from the reported protocol in Ref.~\citenum{iQCC} is referred to as the direct interaction set (DIS). Each element from $\{\mathcal{P}\}$ can be exponentiated with a rotational parameter $\phi_j$ to form a generator $\mathcal{G}_j = \exp\left(-\phi_j\mathcal{P}_j\right)$. After $M$ unique generators are created and applied sequentially onto the circuit, the ansatz is optimized as with any other VQAs.

Subsequent QCC iteration can be applied by folding the generators into the Hamiltonian. That is, folding the associated Pauli operators into the Hamiltonian with the following equation,

\begin{align}
    \hat{H}_{m+1} &= e^{i\phi_m\mathcal{P}_m} \hat{H}_{m}e^{-i\phi_m\mathcal{P}_m} \\
    &= \hat{H}_{m} - \frac{i}{2} \sin{\phi_{m}}\left[\hat{H}_{m}, \mathcal{P}_{m}\right] + \frac{1}{2} (1 - \cos{\phi_{m}}) \left( \mathcal{P}_{m} \hat{H}_{m} \mathcal{P}_{m} - \hat{H}_{m} \right) . \label{eq:folding}
\end{align}

This is recursively applied for each generator in reverse order of the appearance in the circuit such that $\hat{H}_m$ is the Hamiltonian at step $m$ and $\hat{H}_{m+1}$ is the Hamiltonian used for the next QCC step, noting that $\hat{H}_{0}$ is the initial qubit Hamiltonian.

% Problems for VQAs. Barren plateaus, measurements, etc. 
% Already introduced in the introduction but more details here. Can be moved earlier?
The optimization of parameters in VQAs presents well-known challenges. First, ansatz expressivity often comes at the cost of trainability due to the occurrence of barren plateaus~\cite{mcclean2018barren,BarrenPlateau}. The iterative nature of iQCC aims to mitigate this problem by avoiding long depth circuits with many parameters. Still, large expectation value variances can be present in the Hamiltonian~\cite{Measure} when measuring term-by-term or in commuting groups. Therefore, many shots in many different measurement bases are required to obtain accurate results. This additional obstacle in VQAs is commonly known as the measurement problem. A technique to mitigate this problem is to start with a good initial guess for the optimal parameters, to reduce the number of optimizations steps $\lambda$ required for convergence. Several methods to find the best parameter set initialization have been proposed by using Clifford versions of the ansatz (quadratic Clifford expansion~\cite{mitarai2022quadratic}, Clifford circuit annealing~\cite{cheng2022clifford}, and Clifford Ansatz For Quantum Accuracy (CAFQA)~\cite{Ravi2022CAFQA}). The success of Clifford pre-optimization relies on the resulting stabilizer state having non-zero expectation values with the Pauli terms in the cost function. That is, it returns information that can be used. This makes it more likely to be practical for Hamiltonian classes having many non-commuting groups~\cite{cheng2022clifford}.

\section{Clifford-iQCC}\label{sec:clifford_iqcc}

\subsection{Algorithm Overview}\label{sec:general_idea}

% Description of what components are we trying to fix.
Here, we present an algorithm that inherits the advantages of iQCC while avoiding the measurement problem by utilizing Clifford circuits. This is done by observing that the exponential of a Pauli word is composed of only Clifford gates for rotations at $\theta = \pm \pi/2$ when compiled to a circuit using Ref.~\citenum{Time-Evolution}, and that these are also the angles required for one Rotosolve step~\cite{Rotosolve}, employed as an analytical optimizer to minimize the energy function with respect to the rotation angle. As Clifford circuits can be classically simulated~\cite{Clifford}, we can obtain exact optimal parameters and energies for every candidate Pauli word and choose the one that lowers the energy the most at each iteration without utilizing a quantum computer. Each chosen Pauli word is then folded into the Hamiltonian using Eq.~\ref{eq:folding} and the process is repeated until convergence. No quantum computing resources are required to implement this algorithm as it is performed completely classically. In addition, solutions from this quantum-inspired method can be naturally mapped to a quantum circuit to serve as an initial state preparation for subsequent application of quantum algorithms. This process is shown in Algorithm~\ref{alg:clifford_iqcc}, and is described in more details in Section~\ref{sec:details}.

\begin{algorithm}[H]
    \caption{Clifford iQCC algorithm for one iteration}\label{alg:clifford_iqcc}
    \begin{algorithmic}
        \Require $\hat{H}_0$, $E_0$
        \Require $\ket{\psi_0}$ can be represented by a Clifford circuit
        \State $m \gets 0$

        \Do
        \State $m \gets m+1$

        \State COMPUTE $\{\mathcal{P}\}$ from DIS of $\hat{H}_{m-1}$

        \For{$\mathcal{P}_j \in \{\mathcal{P}\}$}
            \For{$\phi \in \{\pi/2, -\pi/2\}$}
                \State $E_{\phi} \gets \bra{\psi_0} e^{i\phi \mathcal{P}_j} \hat{H}_{m-1} e^{-i\phi \mathcal{P}_j} \ket{\psi_0}$
            \EndFor
            \State $E_{\mathcal{P}_j}$, $\phi_{\mathcal{P}_{j}}$ $\gets$ ROTOSOLVE($E_{m-1}, E_{\pi/2}, E_{-\pi/2}$)
        \EndFor
        
        \State SELECT $\mathcal{P}_{j}$ where $E_{min} = \text{MIN}(\{E_{\mathcal{P}_j}\})$
        \State $E_m \gets E_{min}$, $\mathcal{P}_{m} \gets \mathcal{P}_{j}$, $\phi_{m} \gets \phi_{\mathcal{P}_j}$
        \State $\hat{H}_{m} \gets e^{i\phi_m\mathcal{P}_m} \hat{H}_{m-1}e^{-i\phi_m\mathcal{P}_m}$
        
        \doWhile{$|E_m - E_{m-1}| > \epsilon$}
    \end{algorithmic}
\end{algorithm}

\subsection{Implementation details}\label{sec:details}

% Step 1: reference state (HF).
The algorithm begins with the generation of a Hamiltonian in the form of a qubit operator and the corresponding Hartree-Fock reference state. In addition to the JW and BK mappings, we employed for the first time to our knowledge the JKMN mapping~\cite{JKMN} for the QCC ansatz in VQA applications. It was found for this mapping that candidate $\mathcal{P}_{j}$ operators for the QCC ansatz must respect $1 \leq \lvert\mathcal{P}_{j}\rvert \leq N$ to achieve convergence of the total electronic energy (see Appendix~\ref{appendix:jkmn_for_qcc} for more details on this point). In QCC, a quantum mean field (QMF)~\cite{QCC} state is used. This can be initialized to the same as the Hartree-Fock reference, but also can include additional $R_X$ or $R_Z$ rotations. For this algorithm, we require that the QMF state use only Clifford gates. This means that only rotation angles that are multiples of $\pi/2$ are permitted. For the purposes of this work, we restrict ourselves to the Hartree-Fock reference state which has all $R_Z$ angles zero and $R_X$ angles of $0$ or $\pi$, depending on the orbital occupations and the chosen mapping. Other initial reference states can be utilized, provided that the preparation circuit consists exclusively of Clifford gates, similar to the Clifford initialization techniques previously reported in the literature~\cite{mitarai2022quadratic,cheng2022clifford,Ravi2022CAFQA}.

% Step 2: DIS and generators.
After the reference QMF circuit ($\ket{\psi_0}$) and initial qubit Hamiltonian ($\hat{H}_0$) are generated. We first calculate the starting energy $E_0$ using established methods~\cite{QMFener}. The DIS of energy-lowering candidate Pauli words ($\{\mathcal{P}\}$) described in Ref.~\citenum{iQCC}, is then computed from $\hat{H}_0$. For each candidate Pauli word, a circuit primitive corresponding to $\pm\pi/2$ using the CNOT ladder construction of Ref.~\citenum{Time-Evolution} is constructed. Those circuit primitives are also Clifford circuits, as parameterized gates can be expressed according to the following identities:

\begin{align*}
    R_Z(\pi/2) &= HS^{\dagger}HS^{\dagger}H \\
    R_Z(-\pi/2) &= HSHSH \\
    R_X(\pi/2) &= S^{\dagger}HS^{\dagger} \\
    R_X(-\pi/2) &= SHS
\end{align*}

% Step 3: Evaluate generators with Rotosolve.
where $H$ and $S$ are the Hadamard gate and the phase gate $\sqrt{Z}$, respectively. Using a stabilizer emulator, such as Stim~\cite{gidney2021stim}, one then performs an energy evaluation for both parameters $\pm \pi/2$. The Rotosolve algorithms can then be used to calculate the exact minimum, and corresponding rotation angle for every generator. This is done by using Eqs.~\ref{eq:phi_pj}-~\ref{eq:e_pj}.

\begin{align}
     \phi_{\mathcal{P}_j} &= -\frac{\pi}{2} - \mbox{atan2}\left(2 E_{m-1} - E_{\pi/2} -E_{-\pi/2}, E_{\pi/2} - E_{-\pi/2}\right) \label{eq:phi_pj} \\
     A &= \frac{1}{2}\sqrt{\left(2 E_{m-1} - E_{\pi/2} - E_{-\pi/2} \right)^2 + \left(E_{\pi/2} - E_{-\pi/2} \right)^2} \\
     B &= \mbox{atan2}\left(2 E_{m-1} - E_{\pi/2} - E_{-\pi/2}, E_{\pi/2} - E_{-\pi/2}\right) \\
     C &= \frac{1}{2}\left(E_{\pi/2} + E_{-\pi/2}\right) \\
     E_{\mathcal{P}_j} &= A\sin\left(\phi_{\mathcal{P}_j} + B \right) + C \label{eq:e_pj}
\end{align}

% Step 4-5: Select right generator and fold the Hamiltonian.
where $E_{m-1}$ and $E_{\pm \pi/2}$ are the expectation values at the previous iteration, and after adding the generator with parameters $\phi = \pm \pi/2$, respectively. When using a QMF state, one could calculate the expectation values by expanding the Hamiltonian using Eq.~\ref{eq:folding} with $\phi_j=\pm \pi/2$ and calculating the expectation value with the QMF reference using Ref.~\citenum{QMFener}.

The next step is to select the generator $\mathcal{P}_j \in \{\mathcal{P}\}$ that minimizes the energy the most. This differentiates the Clifford iQCC algorithm from the original implementation because no gradient is computed, and the generator having the most significant impact on minimizing the energy is selected from this screening procedure. Finally, we check if the new computed energy $E_{min}$ is converged to a predetermined threshold using the previous iteration expectation value $E_{m-1}$. If not, the algorithm continues with the new expectation value, and the exponentiated Pauli word and parameter are folded into the Hamiltonian by Eq.~\ref{eq:folding}. Another iteration can then be started with the new folded Hamiltonian.

% Unfolding the Hamiltonian into a circuit.
The complete circuit after all iterations can be compiled by applying the exponentiated Pauli words in reverse order (last operator to first) using the optimal angles generated. The expectation value of this circuit with the original qubit Hamiltonian will be equivalent to the minimized expectation value. It should be noted that the parameters generated, although optimal at each step, are not optimal across the whole circuit. Further optimization on a quantum computer after each $\mathcal{P}_m$ is added would provide convergence to a desired accuracy in fewer iterations. This secondary optimization, described in Appendix~\ref{sec:opt_int_gens}, can be performed using the full circuit and the original Hamiltonian. This unfolding into a quantum circuit would be appropriate once the classical simulation becomes bottlenecked by the growth of Hamiltonian terms.

\section{Optimization for Scaling-up the System Size}\label{sec:scaling_up}

As is, our algorithm suffers from the same drawbacks of the algorithm it was inspired from, namely iQCC, described in Section~\ref{sec:qcc}. The Hamiltonian's size becomes intractable as the number of iterations grows, and this is hindering the algorithm ability of handling increasing problem sizes. We show in this section strategies to improve the algorithm's efficiency, which have been necessary to handle the computation of a 40-qubit organometallic complex, mapped to 40 qubits, as shown in Section~\ref{sec:results_catalyst}.

\subsection{Epstein-Nesbet Perturbation Theory}\label{sec:enpt}

An obvious optimization to mitigate the Hamiltonian growth is by achieving faster convergence. A perturbation correction can be applied, while reusing the generator information that is normally thrown away after each iteration. 

Following Ref.~\citenum{ryabinkinPosteriori2021}, we can utilize the full set of DIS generators, and use the previously known optimal parameters for approximating the full state preparation. The perturbative energy is computed by simply summing up the energy differences obtained from each generator. The generator selection in our algorithm uses Algorithm~\ref{alg:clifford_iqcc}, and it implies that the generator selection uses the minimized energy for the whole operator set, therefore the computation of the perturbative term does not involve a significant overhead. However, at a given iteration of the algorithm, this comes at the cost of increasing the equivalent circuit depth since more generators have to be considered. This trade off between Hamiltonian size and circuit depth is likely to depend on the target device architecture, and exploring it is beyond the scope of this article.

\subsection{Generation of anti-commuting generators}\label{sec:gen_anticom_gen}

Past work highlighted the involutory property of qubit generators to mitigate the growth of the Hamiltonian~\cite{ILC}. More recently, follow-up results were published, and they described how to generate a set of involutory linear combination (ILC) of anti-commuting Paulis while respecting a generator ranking procedure~\cite{ryabinkin2023efficient}. This technique was used in this work to mitigate the Hamiltonian growth when scaling up the system size. The process we intend to report here is similar, with the exception that there is a generalization for non-primary generators.

The algorithm begins with a diagonalization of the X-indices in the binary representation of the direct interaction set (DIS). After this step, the primary indices are the columns with a single 1 as defined in Ref.~\citenum{ryabinkin2023efficient} for the matrix $\textbf{M}_{\text{rref}}$. The operators are constructed by alternating $YX$ (or $XY$) in such a way that the number of $Y$ is always odd. After the fact, $Z$s are prepended before $Y$, and $Z$s appended after $X$.

The commutativity of the above operators can be verified by examining the flip-indices through element-wise and summation operations on the bit strings. For example, a three flip-index operator commutes with a two-index operator when zero or two flip-indices match, but not one. It is found to be easier to generate a candidate operator using the above protocol, cycle through the previously generator set, and check that it indeed anti-commutes with each and everyone of them. Occasionally, there are full rows of zeros after performing the diagonalization if the DIS is not of full rank.

\subsection{Restricting the qubit active space}\label{sec:qubit_active_space}

% What is a tableau?
In practice, we used Stim~\cite{gidney2021stim}, which is a "tool for high performance simulation and analysis of quantum stabilizer circuits", to simulate Clifford circuits with the algorithm reported in Ref.~\citenum{Clifford}. This algorithm encodes the quantum state into a \emph{tableau}, which consists of a $\mathcal{O}(4n^2)$ bit matrix representing $n$ stabilizer and $n$ destabilizer generators ($n$ is the number of qubits). These bits can be updated efficiently in polynomial time with a classical computer, as demonstrated by Gottesman and Knill~\cite{gottesman1998heisenberg}. We report here a procedure to reduce the classical overhead of the algorithm by restricting the simulations to a restricted qubit space.

% Summary of the idea, i.e pushing qubit indices in the lowest index qubits.
The idea is that one can use the concepts described in Section~\ref{sec:gen_anticom_gen} to preferentially push the flip indices with the highest decrease in energies to the lowest qubit indices. This is exactly what the diagonalization process accomplishes in the ILC procedure. To construct the Clifford tableau that defines the general transformation, one simply needs to perform the Gaussian elimination over binary field with the same operations applied to the identity matrix~\cite{ryabinkin2023efficient}. We can denote this process as $\left(X|I\right)\rightarrow \left(X^{\prime},T_{xx}\right)$. The corresponding matrix $T_{xx}^{T}$ (where $T_{xx}^{T}$ is the transpose of $T_{xx}$) defines the upper left quadrant of the tableau given in Ref.~\citenum{Clifford}. To obtain the lower right $T_{zz}$ matrix, Gaussian elimination over binary field can be utilized as $\left(T_{xx}|I\right)\rightarrow \left(I,T_{zz}\right)$. The other two quadrants of the tableau (namely $T_{xz}$ and $T_{zx}$) are zero matrices. This completely defines the quantum transformation, encoded in a tableau, that brings the qubit into the form such that the terms that lower the energy the greatest are diagonal or acting on the lower index qubits. After this process, some of the qubits may exhibit zero contribution to the energy gradients, and they can be excluded from the simulation without loss of information.

% Coding the tableau, and getting the circuit transformation (only CNOTs). Switching between reduced and full qubit space.
%The tableau can be instantiated with the python code \lstinline{tableau=stim.Tableau.from_numpy(x2x=T_xxT, z2x=zeros, x2z=zeros, z2z=T_zz)}. Each term is then transformed by \lstinline{new_pword=tableau(pword)} where \lstinline{pword} is one of the Pauli operators in the original Hamiltonian and \lstinline{new_pword} is the Pauli operator in the new Hamiltonian. It is always possible to obtain a circuit that defines this transformation. In Stim, one can use \lstinline{tableau.to_circuit(method="elimination")}.
The resulting transformation is transpiled into a circuit constructed from solely CNOT gates~\cite{harrison2024sierpinski}, so it possible to apply and revert it on a physical quantum device. Therefore, one can always perform a state preparation in the reduced space and then apply the inverse of this circuit followed by the Hartree-Fock state preparation to obtain the transformation back to the original Hamiltonian if desired. This is true in the fault-tolerant era as only Clifford gates determine this transformation and are thought to have low overhead.

\subsection{Efficient energy evaluations}\label{sec:efficient_e_eval}

% Certain properties for efficient energy evaluation.
One can take advantage of certain properties to make the energy evaluation more efficient if the reference state is set as the zero state $\ket{0}^{\otimes N}$. Enforcing this reference state is always possible when the chosen initial state is composed of only Clifford gates, as classical transformations can still be done efficiently.

% Process with the flip indices.
For the energy evaluation, measurement bases that include solely $Z$ and $I$ terms on each qubit contribute to the energy. Likewise, for determining the gradient of a generator with $X$ or $Y$ on a set of indices, it is only necessary to measure $\bra{0} HP \ket{0}$ for a set of generators which can be determined in the following manner. The magnitude of the gradient only depends on those terms with matching positions of $X$ and $Y$ with $Z$ or $I$ on all other qubits. The positions of the qubits with $X$ and $Y$ are known as the \emph{flip} indices~\cite{ryabinkin2023efficient} and are what define the Ising decomposition of Refs.~\cite{iQCC,ryabinkinPosteriori2021}. To screen generators with a substantial gradient, one can simply run through each term in the Hamiltonian, determine its flip indices and calculate the expectation value for a candidate generator with the same flip indices. Each candidate with unique flip indices will have its own sum over the relevant non-zero terms while cycling through the Hamiltonian terms.

% Heisenberg representation.
To determine the optimal angle and energy, the Heisenberg evolution of a Hamiltonian with respect to the candidate generator $P$ can be used, which is,

\begin{equation}\label{eq:heis}
    e^{-iP\phi}He^{iP\phi} = \cos^2{\phi}\ H - \frac{i}{2} \sin{2\phi} [H,P] +\sin^2{\phi}PHP.
\end{equation}

The expectation value of this transformed Hamiltonian with respect to $\ket{0}$ involves measuring the gradient, i.e. the commutator $\bra{0} [H,P] \ket{0}$, and the term $\bra{0} PHP \ket{0}$. The commutator is already accessible after ranking the generator set. $\bra{0} PHP \ket{0}$ can be calculated quickly as Hamiltonian terms with only $Z$ and $I$ operations are non-zero. As the Hamiltonian grows, this becomes an exponentially smaller part of the Hamiltonian. The minimum energy and angle can be obtained by solving Eq.~\ref{eq:heis} for its minimum angle which is given by

\begin{equation}
    \phi_{min} = 2\arctan \left[\dfrac{\sqrt{a^2-2ac+4b^2}-a+c}{2b}\right]
\end{equation}

% Expectation value of a Pauli.
with $b\neq 0$, and where $a=\bra{0} H \ket{0}$, $b=\bra{0} HP \ket{0}$ and $a=\bra{0} PHP \ket{0}$. The expectation value of a Pauli word $\bra{0} \mathcal{P} \ket{0}$ containing only $Z$ and $I$ operations is always 1. Therefore, only the phase change that occurs with the multiplication $HP$ or $PHP$ needs to be accounted for.

\section{Results}\label{sec:results}

\subsection{Computational details}

% Intro and details on the small systems.
Here, we present results for a set of molecular systems. Inspired from Ref.~\citenum{iQCC}, we used a linear chain of hydrogen atoms \ce{H4}, lithium hydride \ce{LiH}, and water \ce{H2O}. The bond distances for \ce{H4} and \ce{LiH} were \qty{1.0}{\angstrom} and \qty{1.7}{\angstrom}, respectively. For the water molecule, the hydrogen and oxygen atom bond distance was set to \qty{1.25}{\angstrom}, while $\angle \ce{HOH} = \ang{107.6}$. All those systems were described in the STO-3G basis set, and the frozen-core approximation was used.

% Using Tangelo and number of qubits.
All calculations were performed using Tangelo~\cite{tangelo}, with PySCF~\cite{pyscf1,pyscf2} as a backend quantum chemistry package to compute the molecular integrals. The fermionic Hamiltonian was transformed into a qubit form, as shown in Eq.~\ref{eq:qubit_hamiltonian}, using the Jordan-Wigner~\cite{jordanUber1928}, Bravyi-Kitaev~\cite{bravyi2002fermionic} or the JKMN~\cite{JKMN} transformations. The number of qubits required for each molecular system are respectively 8, 10 and 12 for \ce{H4}, \ce{LiH}, and \ce{H2O}. The results were compared to their respective Full Configuration Interaction (FCI) solutions, which correspond to the lowest eigenvalue of the Hamiltonian.

\subsection{Results for Small Systems}\label{sec:results_small_systems}

\begin{figure}[h!]
    \centering
    \includegraphics{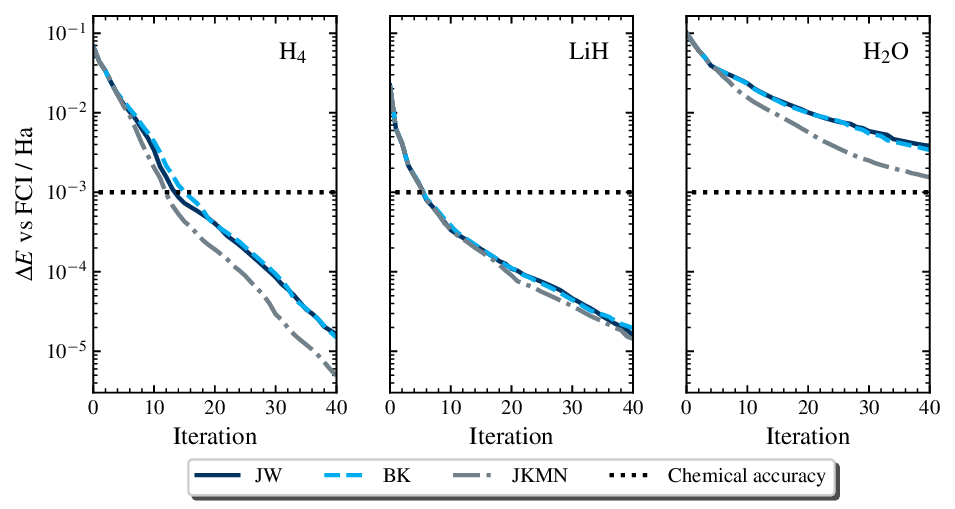}
    \caption{Energy convergence of Clifford iQCC with different qubit mappings. The method is able to output an energy below chemical accuracy for \ce{H4} and \ce{LiH} within 15 iterations, while it would require more than 40 iterations for \ce{H2O}. Similar convergence behavior is observed across the different qubit mappings, although the JKMN one seems to converge more rapidly in some cases.}
    \label{fig:energy_small_systems}
\end{figure}

% Energy convergence on small systems.
As noted in Ref.~\citenum{cheng2022clifford}, the chosen mapping determines where in the full N-qubit Hilbert space the ground state is located. Results for the tests systems in the three mappings discussed earlier are shown in Fig.~\ref{fig:energy_small_systems}. We found that the mappings are generally converging at a similar rate, with the JKMN ones converging slightly faster in some situations. The algorithm reaches chemical accuracy within a few iterations for \ce{H4} and \ce{LiH}. For the \ce{H2O} molecule, chemical accuracy is not reached after 40 iterations (error of about \qty{1.5}{\milli\hartree}), but the curve shape is still exhibiting energy diminution. As illustrated in Fig.~\ref{fig:energy_small_systems}, the accuracy can reach \qty{e-5}{\hartree} after 40 iterations, with no obvious plateaus. This shows that it may be possible to generate a state with an arbitrary energy accuracy.

\begin{figure}[h!]
    \centering
    \includegraphics{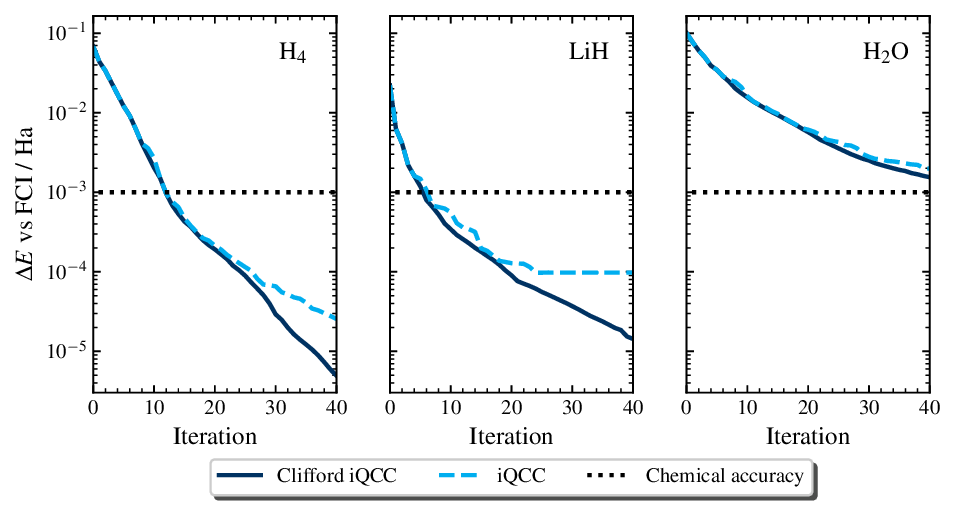}
    \caption{Energy convergence of Clifford iQCC versus the original iQCC implementation. The qubit mapping has been set to JKMN, and a single generator was added at each iteration, which corresponds to $N_g = 1$ in Ref.~\citenum{iQCC}. One of the main difference lies in the generator selection, and it appears to be equivalent in the first iterations of the algorithms. Beyond chemical accuracy, the refinement of the ground state benefits from the Clifford iQCC generator screening method, which selects the generator that minimizes the energy the most.}
    \label{fig:energy_vs_vanilla_small_systems}
\end{figure}

% Faster convergence using energy selection (vs gradient).
In the original implementation of the iQCC, the gradient was used to select generators~\cite{QCC}. To assess if it is worthwhile to perform the Clifford evaluations for all generators, we compare the iQCC energy convergence, which is based on a protocol where the next generator is selected by the highest gradient. The alternative protocol for Clifford iQCC would be to preselect the highest gradient and perform Rotosolve for the selected $\mathcal{P}_j$ only. In Fig.~\ref{fig:energy_vs_vanilla_small_systems}, similar convergence is observed at the beginning of the iterative algorithm, where the gradient and energy changes are expected to be large. When approaching (and beyond) chemical accuracy, gradient magnitudes are much smaller and it is beneficial to actually select the generator that minimize the energy the most.

The key distinction between our implementation and the original approach lies in their intended execution environments: the iQCC is designed for iterative runs on a quantum computer, while the Clifford iQCC is optimized for classical computation. Our approach enables the generation of initial states without being limited by quantum device constraints, such as hardware noise and sampling accuracy. Although the growth in the number of terms in the Hamiltonian remains the primary bottleneck in the workflow, similar to the original iQCC, Clifford iQCC streamlines execution by eliminating the need for quantum measurements.

As stated in the "No Free Lunch" theorem, the main, and possibly insurmountable, issue with this algorithm is the exponential growth in Hamiltonian terms that results from Eq.~\ref{eq:folding}. Considering the time-reversal symmetry of the Hamiltonian, a pair number of $Y$s must be in each Pauli word of Eq.~\ref{eq:qubit_hamiltonian}. Enforcing this condition, as mentionned in Ref.~\citenum{ILC}, reduces the maximum number of possible Hamiltonian terms, denoted by $N_P$, from $4^{n_{\text{qubits}}}$ to Eq.~\ref{eq:max_n_h_terms}.

\begin{equation}\label{eq:max_n_h_terms}
    N_P = \sum_{m=0}^{\lfloor n_{\text{qubits}} / 2 \rfloor} \binom{n_{\text{qubits}}}{2m} 3^{n_{\text{qubits}}-2m}
\end{equation}

\begin{figure}[h!]
    \centering
    \includegraphics{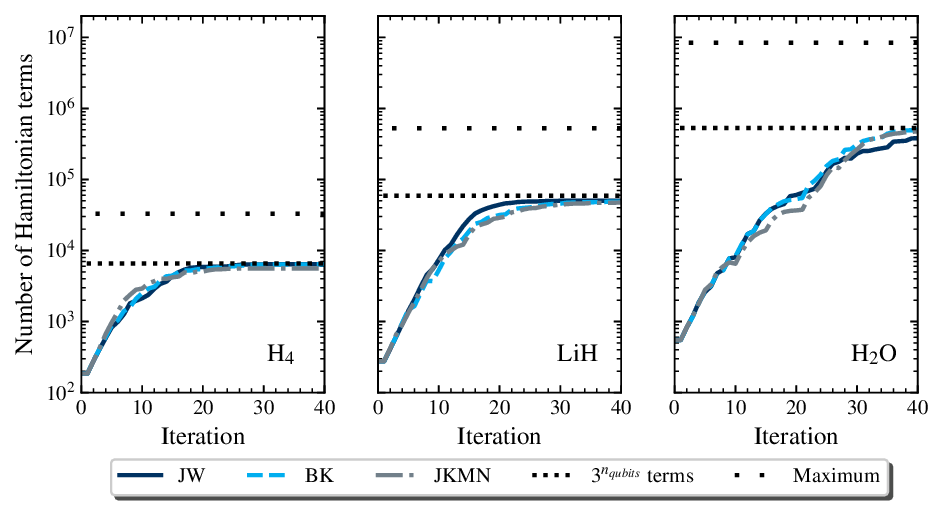}
    \caption{The growth in the number of Hamiltonian terms with respect to the number of Clifford iQCC steps. The number of terms levels off around $3^{n_{\text{qubits}}}$, and it is about one order of magnitude less than the maximum computed with Eq.~\ref{eq:max_n_h_terms}.}
    \label{fig:hterms_small_systems}
\end{figure}

\begin{figure}[h!]
    \centering
    \includegraphics{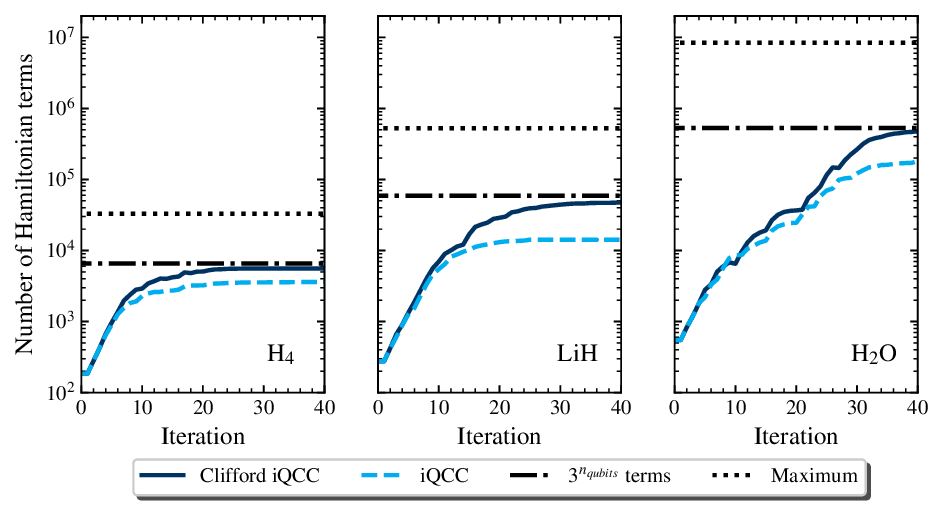}
    \caption{The growth in the number of Hamiltonian terms comparison with the original implementation as in Ref.~\citenum{iQCC}. The qubit mapping has been set to JKMN, and a single generator was added at each iteration, which corresponds to $N_g = 1$. The number of terms for Clifford iQCC is larger than for iQCC, which could be an element of the explanation for the different energy convergence behavior in Fig.~\ref{fig:energy_vs_vanilla_small_systems}.}
    \label{fig:hterms_vs_vanilla_small_systems}
\end{figure}

In practice, the number of Hamiltonian terms in iQCC seems to level off far below the maximum number of possible terms, as jointly stated in Ref.~\citenum{Genin_2022} and illustrated in Figs.~\ref{fig:hterms_small_systems}-\ref{fig:hterms_vs_vanilla_small_systems}. For example, the number of Hamiltonian terms levels off at about \num{6e3}, \num{4.8e4}, and \num{4.5e5} terms for \ce{H4}, \ce{LiH}, and \ce{H2O} respectively. This is generally one order of magnitude below their respective theoretical maximum of about \num{3.3e4}, \num{5.2e5}, and \num{8.4e6}.

The Hamiltonian growth is still expected to be a prohibitive problem, and this affects the computational cost for the classical computer. This is reflected in the requirement to keep the Hamiltonian in memory, and performing Hamiltonian transformation on this data as depicted in Eq.~\ref{eq:folding}.

\begin{figure}[h!]
    \centering
    \includegraphics{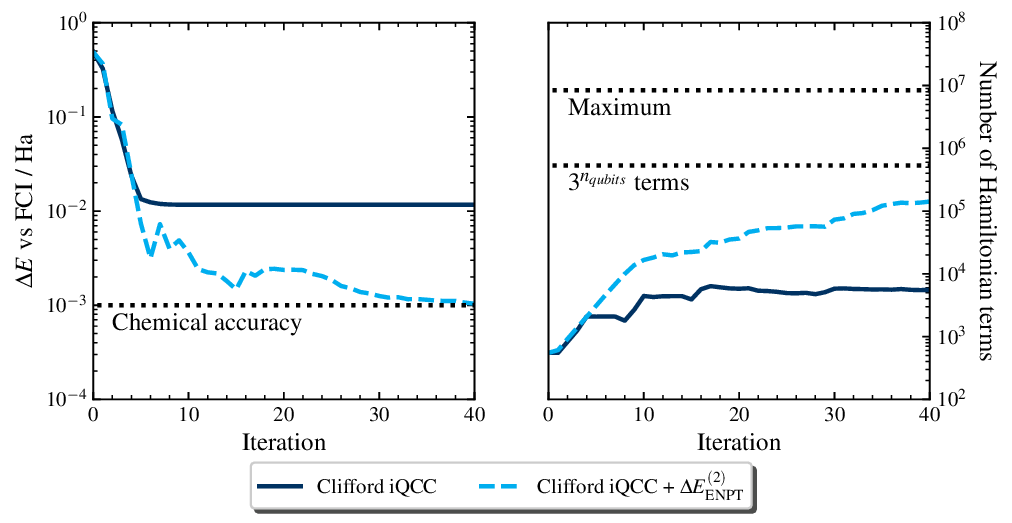}
    \caption{Clifford iQCC results for the stretched \ce{H2O} molecule. The addition of the perturbative terms at each iteration makes the energy convergence more rapid. Without the correction, the algorithm shows signs of being stuck in a barren plateau, and this is not the case after considering the perturbative terms.}
    \label{fig:stretched_h2o}
\end{figure}

We also explored the use of perturbative corrections, described in Section~\ref{sec:enpt}, and found that this not only generates a reasonable approximation to the energy, but the energy of the resulting circuit has good agreement with the perturbative energy itself and has good overlap with the desired state. This is illustrated by the results of a stretched configuration of the \ce{H2O} molecule, where the OH bond is set at \qty{2.35}{\angstrom} (Fig.~\ref{fig:stretched_h2o}). Similar to what was done in Ref.~\citenum{iQCC}, this system is studied to assess if the method is able to treat strong correlation. The Clifford iQCC without perturbative corrections does converge to a solution, but it is still far from chemical accuracy even after 40 iterations. Considering the correction results in an energy evaluation brings the solution significantly closer to the FCI result in this case.

As a side note, we have heuristically found that when two generators have the same $\phi_m$ and $E_{\mathcal{P}_j}$, it is better to keep both generators but utilize an angle of $\phi_m/\sqrt{2}$ for each. This is especially true during early iterations of the iterative process.

\subsection{Results for an organometallic complex}\label{sec:results_catalyst}

\begin{figure}[h!]
    \centering
    \includegraphics{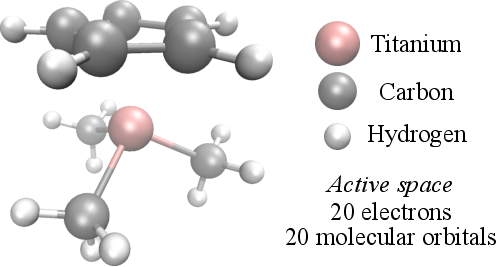}
    \caption{Molecular system used to test the algorithm optimizations. The titanium-based catalyst (\ce{Ti(C5H5)(CH3)3}) ground state calculation has been restricted to a (20, 20) active space.}
    \label{fig:catalyst}
\end{figure}

We perform the ground state calculation, on a personal laptop, of a 20 electrons in 20 orbital active space (40 qubit problem) of a Ti-based catalyst \ce{Ti(C5H5)(CH3)3}, as shown in Fig.~\ref{fig:catalyst}, in the STO-3G basis. We perform 20 Clifford iQCC iterations using Tangelo along with an interfaced in-house \verb!C++! code. 

\begin{figure}[h!]
    \centering
    \includegraphics{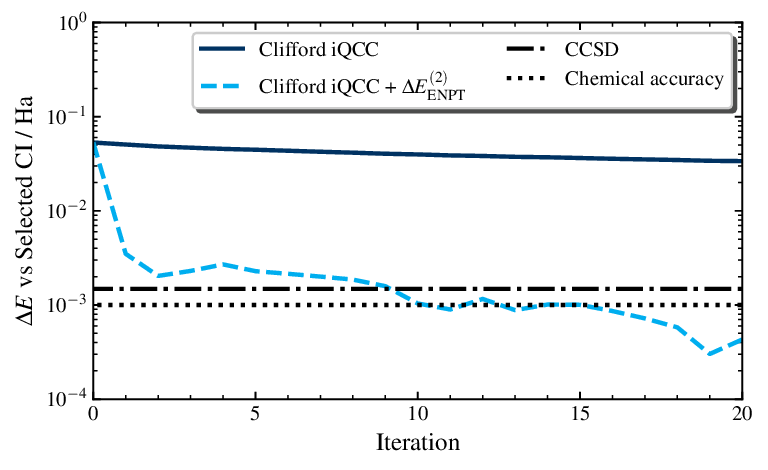}
    \caption{Energy convergence of a (20, 20) system, corresponding to a 40-qubit calculation in the occupancy qubit mapping. This calculation has been done on a personal laptop after the optimization mentioned in Section~\ref{sec:scaling_up} were implemented. As the FCI solution is not accessible for this system, the comparison is made versus the Selected Configuration Interaction (SCI) method. The Clifford iQCC convergences very slowly, and still outputs a significant energy error after 20 iterations. Adding the perturbative correction terms described in Section~\ref{sec:enpt} makes the algorithm converged to a reasonable solution after few iterations, and it reaches an SCI-like state after 10 iterations.}
    \label{fig:energy_catalyst}
\end{figure}

As can be seen in Fig.~\ref{fig:energy_catalyst}, the algorithm including perturbative corrections outputs results closer to the selected configuration interaction (SCI) than coupled cluster singles and doubles (CCSD) in fewer iterations. We observed that the number of qubits the generators act on is restricted the lowest 12 for the first 20 iterations. This means that the optimization can be performed on only the first 12 qubits using the method described in Section~\ref{sec:qubit_active_space}, thus increasing the computational efficiency. However, the computation of the ENPT correction terms depend on all the qubit indices. For the implementation simplicity, we kept track of the eigenvalues on all qubit indices. For future work, it would be relevant to implement a technique that can efficiently generate the perturbation values after performing an optimization on a smaller qubit active space. This should be possible as one only needs to determine how the first $N_{active}$ qubits transform. One could obtain the gradient contributions for the flip-indices assuming all $X$ the candidate generator can always be taken to be $Y$ on the first flip indice and $X$ on the others.

\section{Discussion}\label{sec:discussion}

% Restate the state preparation goal.
Algorithms for state preparation, in the context of quantum algorithms, aim at preparing quantum states where their quality is \textit{neither too high nor too low}~\cite{fomichev2023initial}. This implies that the generated state should be significantly improvable, while also being close enough to the target state to ensure a high probability of success. 

\begin{figure}[h!]
    \centering
    \includegraphics{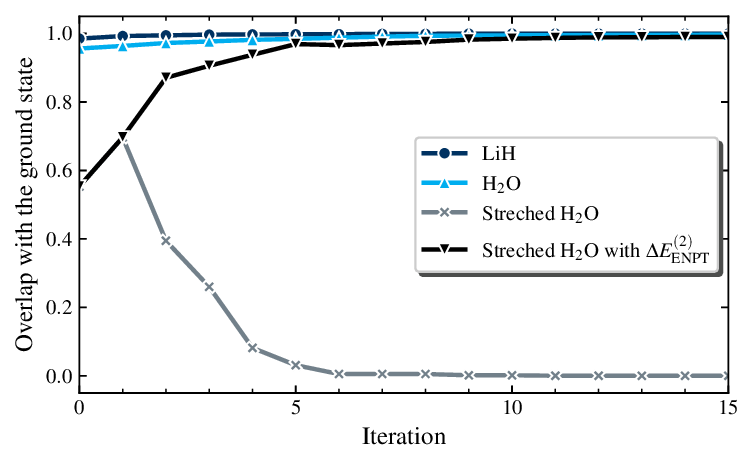}
    \caption{Inner products of the lowest-eigenvalue eigenstate of the Hamiltonian, for several Clifford iQCC runs.}
    \label{fig:overlaps}
\end{figure}

This task should be accomplished within the available classical computational resources and must produce states with a significant overlap with the target state. If the overlap is negligible, the probability of successful QPE decreases, increasing the likelihood that the initial state will project onto an undesired state. Consequently, additional QPE runs will be required to correct this result, increasing the quantum computational cost. Fig.~\ref{fig:overlaps} illustrates the overlap computation of Clifford iQCC runs with the eigenstate corresponding to the lowest eigenvalue of the diagonalized Hamiltonian of Section~\ref{sec:results_small_systems}.

For \ce{LiH} and \ce{H2O}, in their relaxed geometry form, the Hartree-Fock state is already an acceptable initial state. Although, the Clifford iQCC algorithm is able to improve the overlap towards the desired state, which is the ground state in this case. For the stretched \ce{H2O}, the Hartree-Fock state is significantly improved when including the perturbative terms in the Clifford iQCC protocol. Including these terms makes the circuit longer, due to the inclusion of primitives derived from the Pauli words within the DIS procedure, which are usually discarded after each iteration. If these terms are omitted, the algorithm tends to converge toward a different high-energy state. Although the exact reason for this discrepancy remains unknown, this finding highlights the relevance of perturbative terms in multireference systems. Furthermore, these results suggest that the Clifford iQCC can be utilized to produce states with arbitrary overlap.

% Mitigating the Hamiltonian exponential growth.
The algorithm still suffers from an exponential growth of the Hamiltonian, thus hindering its scalability. There are a few avenues that one may use to attenuate this growth of terms. Firstly, most of the Pauli words do not have support on the full Hamiltonian. It may be possible to restrict the DIS such that the rank of candidate Pauli words must be smaller than a particular value. In this case, one can compress the Hamiltonian by evaluating the expectation values of qubits not involved in the candidate Pauli word $\mathcal{P}_{m}$. With a Hartree-Fock reference state, any Hamiltonian term that measures the uninvolved qubits in the Pauli $X$ or $Y$ operator basis will have a zero expectation value, and so can be efficiently removed from the calculation. Secondly, another technique is to explore the possibility of adding multiple generators on the final step. As long as the added Pauli words commute, the order in which you add the generator does not affect the resulting circuit nor the transformed qubit Hamiltonian. Finally, the ILC procedure, described in Section~\ref{sec:gen_anticom_gen}, could also be performed as a pre-processing step before initializing Clifford iQCC as it is also classical pre-processing algorithm to attenuate qubit Hamiltonian growth during the initial steps.

\section{Conclusion}\label{sec:conclusion}

We have highlighted a new a variant of iQCC that is a quantum-inspired classical algorithm capable of minimizing energy while also generating a quantum circuit to simulate the corresponding statevector. This is achieved by evaluating the exponentiated Pauli word circuits at rotations that can be expressed using only Clifford gates. The exponentiated Pauli word is then folded into the Hamiltonian before the next iteration. However, unlike the original iterative QCC algorithm, the reference state can be any Clifford circuit as opposed to being restricted to the QMF form.

In summary, at each iteration the number of parameters remains small, and the number of Pauli terms grows exponentially. This exponential growth is normally detrimental for deploying iQCC on a quantum computer, but can be advantageous for Clifford pre-optimization on a classical computer due to the increase in non-commuting (and non-zero) terms. Fortunately, when the final optimization across the whole circuit is performed on a quantum computer, this can be done with the full circuit and the original qubit Hamiltonian to circumvent measuring an exponential number of terms on hardware.

For future work, in addition to what have been discussed in the discussion, it may be possible to solve for optimal angles and energies using other generators that have a small number of unique eigenvalues~\cite{PSR1, PSR2, QEB} using only Clifford circuits. These include generators of UCC type~\cite{PSR1}. It also may be possible to optimize multiple candidate $\mathcal{P}_m$ at once using Fraxis as it utilizes $\pi/2$ rotation angles~\cite{fraxis}.

Whether this algorithm is applicable to fault-tolerant quantum computing is an open problem, but following the arguments of Ref.~\citenum{cheng2022clifford}, Clifford based initialization is particularly well suited for chemical system use-cases. As there are a large number of non-commuting Pauli strings in a chemical Hamiltonian, Clifford plateaus should not be expected to appear as quickly as for other problems. Although the Hamiltonian grows exponentially, it may be enough to run a few iterations of Clifford iQCC such that phase estimation or adiabatic state preparation succeeds with high probability.

\section{Acknowledgements}
This work was supported as part of a collaboration between Dow and SandboxAQ.

%\bibliography{references}

\begin{footnotesize}
\bibliographystyle{unsrt.bst} % abbrvnat or unsrt
\textnormal{\bibliography{references.bib}}
\end{footnotesize}

\newpage
\appendix

\section{JKMN mapping for QCC}\label{appendix:jkmn_for_qcc}

The JKMN mapping, as described in Reference~\citenum{JKMN}, provides a recipe to generate a Fermion to qubit mapping that is optimal in average Pauli weight of $\log_3(2N)$ through Majorana modes defined as follows:

\begin{equation}
\gamma_{2i} =  \hat{c}_i^\dagger + \hat{c}_i, \quad  \gamma_{2i+i} =  i(\hat{c}_i^\dagger - \hat{c}_i)  
\end{equation}

for spin orbital $i=1,2...N$, and where $\hat{c}_i$ and $\hat{c}^\dagger_i$ are the annihilation and creation operators respectively. This work does not however describe how to allocate these Majorana such that the mapping is amenable to the qubit coupled cluster (QCC) ansatz. For these methods to be compatible we construct a viable vacuum state, and ensure that all occupation operators result in qubit operators with only $Z$ terms. 

The process as implemented in Tangelo~\cite{tangelo} is described by the following recipe.

\begin{enumerate}
    \item Add leaf nodes up to $2N+1$ where $N$ is the number of spin orbitals from left to right in lexographical order as shown in Figure 1 of reference~\citenum{JKMN}. 
    \item Discard the furthest right node corresponding to all $Z$ operators.
    \item Allocate the $2N$ Majorana operators $\gamma_i$ in order from left to right.
    \item Calculate the occupation number operator $\hat{N}_i=\frac{1}{2}(1-i\gamma_{2i}\gamma_{2i+1})$ for all spin orbitals $i$.
    \item Identify which qubits $\{Q\}$ result in an $X$ operator instead of a $Z$ operator and create the unitary transformation $\prod_{j\in \{Q\}}H_j$.
    \item Apply the transformation to each Majorana operator which essentially swaps the labelling.
    \item Each pair $\gamma_{2i}, \gamma_{2i+1}$ will have matching $X$ or $Y$ operations except for 1 qubit. If $Y$ is on $\gamma_{2i}$, change $\gamma_{2i}\rightarrow -\gamma_{2i}$, to ensure that the number operator is $0$ for each qubit. 
\end{enumerate}

To generate the initial state, the occupation vector which $\{O\}$ details the occupied orbitals is used. An excitation operator is determined if $\prod_{i\in\{O\}}\gamma_{2i}=\prod_{i\in\{O\}} \left(\hat{c}_i^{\dagger}+\hat{c}_i\right)$, and the resulting qubit operator signifies which qubits to apply $X$ or $Y$ gates to. This procedure works as $\hat{c}_i$ on the vacuum state is $0$.

\section{Optimizing interior generators}\label{sec:opt_int_gens}

It is possible to optimize interior generators and angles using Clifford circuits. To re-optimize $\phi_{m}$ for interior generator $\mathcal{G}_m$ selected at the $m$th iteration, one creates a qubit operator $Q_{m+1}=\sum_{q} c_q P_q$ that represents the action of the last $m+1$ to $M$ steps as

\begin{equation}
    Q_{m+1} = \prod_{j=m+1}^{M}\left(\cos\left[\phi_j/2\right]-i \sin \left[\phi_j/2\right]\mathcal{P}_j\right),
\end{equation}

The folded Hamiltonian for the first $m-1$ steps (denoted $H_{m-1}=\sum_{j}h_j P_j$) is also computed, which involves using Eq.~\ref{eq:folding} for the first $m-1$ steps.

\begin{figure}[h!]
    \centering
    \begin{quantikz}
        \lstick{\ket{0}} &  & \gate{H} & \ctrl{1} & \qw &\octrl{1} &\qw &\octrl{1} &\\
        \lstick{\ket{\psi_0}} & \qwbundle{N} & & \gate{P_{q}} & \qw & \gate{P_{q^{\prime}}} &\gate{\mathcal{G}_{m}} & \gate{P_j} &
    \end{quantikz}
    \caption{Quantum circuit to optimize interiors generators. $\ket{\psi_0}$ is the initial state as defined in Algorithm~\ref{alg:clifford_iqcc}.}
    \label{fig:reopt_int_gens}
\end{figure}
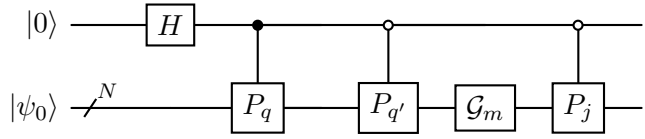

The state $\ket{\psi}= \frac{1}{\sqrt{2}} \left( \ket{0}  \otimes P_j \mathcal{G}_{m}P_{q^{\prime}}\ket{\psi_0} + \ket{1} \otimes \mathcal{G}_{m}P_{q} \ket{\psi_0} \right)$ is then produced using the circuit shown in Fig.~\ref{fig:reopt_int_gens}. The parameterized circuit is evaluated at $\phi=0, \pm \pi/2$ for the selected interior generator $\mathcal{G}_m=\exp\left[-\phi \mathcal{P}_m\right]$ to select the angle $\phi$ that minimizes the energy. For each combination $\{ c_q P_q, \, c_{q^{\prime}}P_{q^{\prime}}, h_jP_j \}$, the expectation value of $X+iY=\left|1\right>\left<0\right|$ of the top register, denoted $E_{qq^{\prime}j}(\phi)$, is computed. The full expectation value is then calculated as 

\begin{equation}
    E(\phi)=\sum_{q q^{\prime} j} c_q c_{q^{\prime}}^{*} h_j E_{qq^{\prime} j}(\phi)
\end{equation}
 
This state preparation only uses Clifford gates because controlled-Pauli operators are Clifford gates, and the generator $\mathcal{G}_{m}$ is evaluated at $\phi \in \{0, -\pi/2, \pi/2 \}$. The above protocol has been verified, but its utility in the context of state preparation has yet to be examined. All parameters could be optimized using multiple Rotosolve sweeps instead of the single sweep we use in this manuscript. This opens up some research questions on the theme of balancing classical and quantum computing resources, for e.g. if swapping generators in place could lead to a better state, or generator ordering have an impact on the generated state overlap with the ground state.

\section{Optimization example}\label{sec:examples}

To illustrate the optimisation procedure described in Section~\ref{sec:scaling_up}, we are showcasing an example which consists of a linear \ce{H4} with \qty{2}{\angstrom} separation in a minimal basis set (STO-3G). This translates into an 8-qubit problem, and a similarity transformation has been applied so that the Hartree-Fock state preparation is included in the Hamiltonian (in other words the reference state is the zero state $\ket{0}^{\otimes 8}$). 

% Generation of ILC generators.
The ILC generation procedure, outlined in Section~\ref{sec:gen_anticom_gen}, results in the identification of 15 energy-lowering Pauli words. When encoded into the binary representation, the matrix

\begin{equation}
\left[\begin{array}{ccccccccccccccc}
     1 & 0 & 0 & 0 & 0 & 0 & 0 & 0 & 0 & 0 & 1 & 1 & 0 & 0 & 1 \\
     0 & 1 & 0 & 0 & 0 & 0 & 0 & 1 & 1 & 0 & 1 & 1 & 1 & 0 & 1 \\
     0 & 0 & 1 & 0 & 0 & 0 & 0 & 0 & 1 & 0 & 1 & 0 & 0 & 0 & 1 \\
     0 & 0 & 0 & 1 & 0 & 0 & 0 & 1 & 0 & 0 & 1 & 0 & 0 & 0 & 1 \\
     0 & 0 & 0 & 0 & 1 & 0 & 0 & 0 & 0 & 0 & 1 & 0 & 1 & 0 & 1 \\
     0 & 0 & 0 & 0 & 0 & 1 & 0 & 0 & 0 & 0 & 1 & 0 & 0 & 1 & 1\\
     0 & 0 & 0 & 0 & 0 & 0 & 1 & 0 & 0 & 0 & 1 & 0 & 0 & 1 & 0 \\
     0 & 0 & 0 & 0 & 0 & 0 & 0 & 0 & 0 & 1 & 0 & 0 & 0 & 1 & 1 \\
\end{array}\right]
\end{equation}

can be found. In this example, the eighth column, i.e. $(0,1,0,1,0,0,0,0)^T$, would result in $$Z_0 X_1 Y_3 Z_4 Z_5 Z_6 Z_7.$$ The last column, i.e. $(1,1,1,1,1,1,0,1)^T$, would correspond to $$X_0 Y_1 X_2 Y_3 X_4 Y_5 Z_6 X_7.$$ Here we used a zero-based indexing scheme for the qubit labels. After performing Gaussian elimination over binary field, the resulting matrix is

\begin{equation}\label{eq:diagdis}
\left[\begin{array}{cccccccccccccc}
     1 & 0 & 0 & 0 & 0 & 0 & 0 & 1 & 1 & 1 & 0 & 0 & 0 & 0 \\
     0 & 1 & 0 & 0 & 1 & 1 & 0 & 1 & 0 & 1 & 0 & 0 & 0 & 1 \\
     0 & 0 & 1 & 0 & 1 & 1 & 1 & 0 & 1 & 0 & 0 & 0 & 1 & 0 \\
     0 & 0 & 0 & 1 & 1 & 0 & 1 & 1 & 1 & 1 & 0 & 1 & 1 & 0 \\
     0 & 0 & 0 & 0 & 0 & 0 & 0 & 0 & 0 & 0 & 1 & 1 & 1 & 1 \\
     0 & 0 & 0 & 0 & 0 & 0 & 0 & 0 & 0 & 0 & 0 & 0 & 0 & 0\\
     0 & 0 & 0 & 0 & 0 & 0 & 0 & 0 & 0 & 0 & 0 & 0 & 0 & 0 \\
     0 & 0 & 0 & 0 & 0 & 0 & 0 & 0 & 0 & 0 & 0 & 0 & 0 & 0 \\
\end{array}\right]
\end{equation}

where one can read the generators in this diagonalized space, in order of column apparition, as

\begin{equation}
    \begin{array}{c|c|c|c|c|c|c}
       Y_0 & Y_1 & Y_2  & Y_3 & Y_4 & Y_1X_2Y_3 & Y_1 X_2  \\ \hline
       Y_2X_3 & Y_0X_1Y_3 & Y_0X_2Y_3 & Y_4 & Y_3X_4 & Y_2 X_3 Y_4 &  Y_2X_4 
    \end{array}
    \label{eq:single_qubit_gen}
\end{equation}

We can also note that the three highest index qubits do not exhibit an energy gradient. The fifth qubit does not contribute to the energy until ten other operators have been utilized. In consequence, most of the correlation can be presumed to be in the four lowest index qubits. The terms in the Hamiltonian can be transformed so that the DIS is defined by Eq.~\ref{eq:diagdis} by applying the tableau to each Pauli operator in the Hamiltonian.

After, applying the corresponding Gaussian elimination steps to identity matrices, one can find the tableau that defines the transformation of the qubit operator that defines this qubit-active space,

\begin{equation}
  T_{xx}  = \left[\begin{array}{cccccccc}
  0 & 0 & 0 & 0 & 0 & 1 & 0 & 0 \\
  0 & 0 & 0 & 0 & 0 & 0 & 1 & 0 \\
  0 & 0 & 0 & 0 & 1 & 1 & 0 & 0 \\
  0 & 0 & 1 & 1 & 1 & 0 & 1 & 0 \\
  0 & 0 & 0 & 0 & 0 & 0 & 0 & 1 \\
  1 & 0 & 0 & 1 & 0 & 1 & 1 & 1 \\
  0 & 0 & 0 & 1 & 1 & 0 & 0 & 1 \\
  1 & 1 & 0 & 1 & 1 & 1 & 1 & 1 \\
    \end{array}\right]
  \label{eq: Txx}  
\end{equation}
\begin{equation}
  T_{zz} = \left[\begin{array}{cccccccc}
  0 & 0 & 0 & 0 & 0 & 1 & 0 & 0 \\
  0 & 0 & 0 & 0 & 0 & 0 & 1 & 0 \\
  0 & 0 & 0 & 0 & 1 & 1 & 0 & 0 \\
  0 & 0 & 1 & 1 & 1 & 0 & 1 & 0 \\
  0 & 0 & 0 & 0 & 0 & 0 & 0 & 1 \\
  1 & 0 & 0 & 1 & 0 & 1 & 1 & 1 \\
  0 & 0 & 0 & 1 & 1 & 0 & 0 & 1 \\
  1 & 1 & 0 & 1 & 1 & 1 & 1 & 1 \\
    \end{array}\right]
  \label{eq: Tzz}    
\end{equation}

and perform the simulation in a reduced qubit-active space using the CNOT transformation circuit as described in Section~\ref{sec:qubit_active_space}.

\section{Hardware efficient ansatz}\label{appendix:hea}

During the numerical simulations described in Section~\ref{sec:scaling_up}, it was observed that the concepts can also be used to construct a customizable circuit ansatz. Its form consists of alternating layers of single qubit rotation gates and entangling gates, traditionally categorized as a hardware efficient ansatz (HEA). The procedure is outlined below: 

\begin{enumerate}
    \item Diagonalize the flip indices of the DIS
    \item Determine the generators which map to single qubits by performing Gaussian elimination over the binary fields, and generate the corresponding Clifford tableau. For example $Y_0,Y_1, Y_2,Y_3,Y_4$ in Eq.~\ref{eq:single_qubit_gen}
    \item Define a transformation of the Hamiltonian using only the selected single-qubit generators.
    \item Obtain the DIS using the new qubit Hamiltonian and remove the already utilized single-qubit generators from the DIS.
    \item Repeat until all desired generators are utilized.
\end{enumerate}

The circuits that perform the transformation do not respect the hardware connectivity so one may need to utilize CNOT decomposition methods~\cite{winder2023architecture,van2023towards,murphy2023global}. The standard hardware efficient ansatz has difficulty converging~\cite{bittel2021training,anschuetz2022quantum,mcclean2018barren} while this QCC hardware efficient ansatz should have no difficulty as the gradients are all large. This is what the optimization process called CHEM~\cite{sun2024toward} does, but it requires a binary optimization problem to be utilized. As the first layer of single qubit gates is trivial to simulate, and is a qubit mean field type~\cite{QCC}, it is more convenient to optimize them separately. A set of these terms could then be used to dress the Hamiltonian to expose a greater number of DIS generators that could be utilized to extend the hardware efficient ansatz.

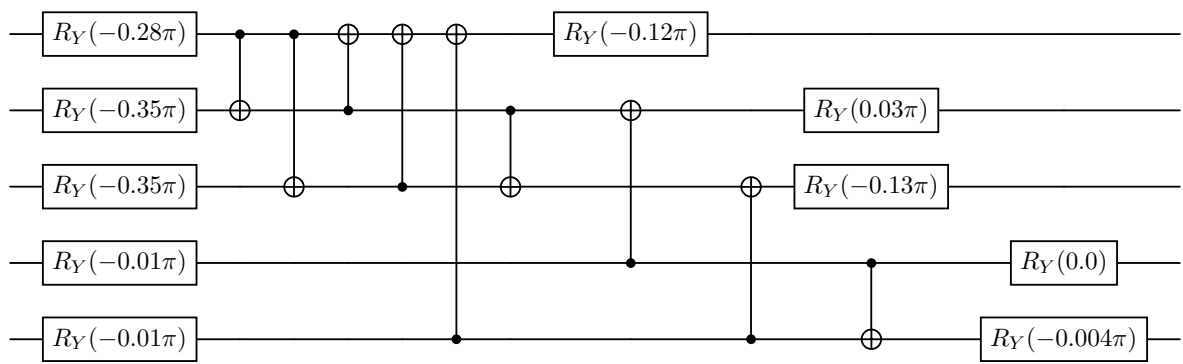
\begin{figure}[h!]
    \centering
    \begin{adjustbox}{width=0.9\textwidth}
    \begin{quantikz}
        & \gate{R_Y(-0.28\pi)} & \ctrl{1} & \ctrl{2}& \targ{} & \targ{} & \targ{} & & \gate{R_Y(-0.12\pi)} & & & & \\
        & \gate{R_Y(-0.35\pi)} & \targ{} & & \ctrl{-1} & & & \ctrl{1} & \targ{} & & \gate{R_Y(0.03\pi)} & & \\
        & \gate{R_Y(-0.35\pi)} & & \targ{} & & \ctrl{-2} & & \targ{} & & \targ{} & \gate{R_Y(-0.13\pi)} &  & \\
        & \gate{R_Y(-0.01\pi)} & & & & & & & \ctrl{-2} & & \ctrl{1} & \gate{R_Y(0.0)} & \\
        & \gate{R_Y(-0.01\pi)} & & & & & \ctrl{-4} & & & \ctrl{-2} & \targ{} & \gate{R_Y(-0.004\pi)} &
    \end{quantikz}
    \end{adjustbox}
    \caption{The hardware efficient ansatz determined by successive diagonalization of the flip indices for linear \ce{H4} in a STO-3G basis.}
    \label{fig:hea}
\end{figure}

As a non-trivial example we are showcasing the linear \ce{H4} molecule in a minimal basis set (STO-3G). Results after diagonalization of the flip indices are shown in Eq.~\ref{eq:diagdis}, and the single-qubit generators are corresponding to columns 1, 2, 3, 4, and 11. The procedure using two layers of this QCC-based HEA produces a state having an error of 9 mH on the molecular energy, with a circuit of only depth 9. The circuit with the optimized parameters is shown in Fig.~\ref{fig:hea}.

%\end{linenumbers}
\end{doublespacing}

\end{document}